\documentstyle [prl,multicol,aps,epsfig]{revtex}
\begin{document}
\title{Effect of resonances on the transport properties of two-dimensional
disordered systems}

\author{Brahim Elattari$^{1,2}$, Tsampikos Kottos$^3$\\
$^1$ Departamento de F\'\i sica Te\'orica de la Materia Condensada,
Universidad Aut\'onoma de Madrid, \\
28049 Madrid, Spain.\\
$^2$ Universit\'e Choua\"\i b Doukkali, Fauclt\'e des Sciences,
El Jadida, Morocco\\
$^3$ Max-Planck-Institut f\"ur Str\"omungsforschung \\
37073 G\"ottingen, Germany}
\vspace{18pt}
\maketitle

\begin{abstract}
We study both analytically and numerically how the electronic structure and
the transport properties of a two-dimensional disordered system are modified
in the presence of resonances. The energy dependence of the density of states
and the localization length at different resonance energies and strengths of
coupling between resonances and random states are determined. The results show,
that at energy equals to the resonance energy there is an enhancement in the
density of states. In contrast, the localization length remains unaffected from
the presence of the resonances and is similar to the one of the standard
Anderson model. Finally, we calculate the diffusion constant as a function of
energy and we reveal interesting analogies with experimental results on light
scattering in the presence of Mie resonances.
\end{abstract}

\hspace{2.5 cm}  PACS: 71.55.Jv, 72.15.Rn

\begin{multicols}{2}

The study of waves propagating in disordered lattices, during the past years
has attracted much attention and many interesting phenomena have been, more
or less, well understood. However, until recently the interest of the
physics community was primarily focusing on quantum, i.e. electronic waves.
It was the observation of the coherent backscattering effect in classical
wave systems \cite{ping}, the analogous effect to weak localization in the
electronic case, which triggered a burst of interest in further studies
of scattering disordered classical wave systems. The question of localization
of classical waves has attracted attention for two reasons. First, the
properties of classical waves such as light waves, microwaves, and acoustic
waves in random media are of fundamental interest for their own sake. Second,
classical waves can serve as a model system for testing the theory of Anderson
localization of electrons experimentally in a clean way, without the
complication of strong inelastic scattering and other effects of electron-
electron and electron-phonon interaction. Existing theories predict the
localization of classical waves under certain circumstances \cite{ping} leading
thus to the conclusion that the analogy between quantum and classical waves
works reasonably well. However, there is no conclusive experimental evidence,
and thus the full correspondence is not yet established beyond any doubt (a
thorough discussion may be found in Refs. \cite{IMRY,GGW97}).

The present paper has been motivated by an experimental work, performed
several years ago \cite{ATLT91} on the scattering of light by a disordered
medium in the presence of Mie resonances. A low concentration of Mie
resonances leads to a strong change in the transport properties of the system,
reflected in a strong reduction of the transport velocity, or equivalently,
of the diffusion constant $D(E)$, near the resonance energies. With
increasing concentration of resonant scatterers, the dip in $D(E)$ widens
and becomes less deep. Qualitatively speaking, the transport velocity is
reduced because a lot of energy is temporarily stored inside the resonance
or equivalently the wave spends a lot of time (dwell time) inside the Mie
resonances.

To understand this surprising result, different theoretical approaches based
on different considerations were developed \cite{ATLT91}, \cite{busch},
\cite{EKW}. However, there are controversies as to whether this can be applied
to the quantum case, i.e., electronic waves. In fact, at low concentration of
Mie resonances \cite{ATLT91}, in the spirit of the coherent--potential
approximation (CPA) \cite{ping}, it was shown that, unlike electronic systems,
the diffusion constant $D(E)$ for classical waves decreases sharply close to
the resonance energies. This result was viewed as due to the different Ward
identity caused by an energy--dependent scattering potential in the classical
case. An extension of the CPA to the strong concentration limit \cite{busch},
the so called coated CPA, shows that the effect of the Mie resonances decreases
at this regime. On the other hand, in a recent work \cite{EKW}, it was shown
that an extension of the Random Matrix Theory (RMT)\cite{rmt}, capable of
accounting for the presence of resonance scattering, is able to explain the
obtained experimental result and generalize it to both quantum and classical
waves, although the corresponding Ward identities are different \cite{EKW1}.

In the present paper, we extend our previous analysis on the effect of
resonances on electrons in a one-dimensional disordered medium \cite{EK99}, to
higher dimensions. The quantities that will monopolize our interest, will be
the density of states (DOS) $\rho(E)$ and the localization length
$l_{\infty}(E)$.
In the one-dimensional case \cite{EK99}, at low concentration and weak coupling
both these quantities are affected strongly by the presence of resonances.
The DOS exhibits a Lorentzian peak at the resonance energy and the localization
length is drastically decreased except at the resonance energy which remains
essentially the same with the one corresponding to the standard Anderson model.
We will show that in higher dimensions, the density of states exhibits the same
resonant enhancement while the localization length remains unaffected in
contrast
to the one-dimensional case. Finally, by making use of the well known results
for
quasi-one-dimensional systems we will compute the diffusion constant $D(E)$ and
identify similarities with the corresponding problem of light scattering in the
presence of Mie resonances.

The mathematical model we consider is the tight-binding Anderson Hamiltonian
on a two-dimensional lattice,
\begin{equation}
\label{andham}
H=\sum_{l} \epsilon_{l}|l><l| +\sum_{l,l'}t_{ll'}|l><l'|
\end{equation}
where $(l)$ denote the sites of a two-dimensional lattice $L\times M$. The
local site energies $\epsilon_{l}$ are taken independently at random within
the interval $[-W/2,W/2]$. At some randomly chosen sites $k$, ($m$ in total),
the energies $\epsilon_{k}$ are taken to be fixed i.e. $\epsilon_{k} = E_r$.
These energies correspond to the resonance energies. To simplify the problem,
we assume that all resonances have the same energy $E_r$. The hopping matrix
elements $t_{ll'}$ are restricted to nearest neighbors. When the coupling is
between random states ($N$ in total), $t_{ll'}$ are constant taken to be the
unit of energy while the coupling between a resonant state and another state
$t_{ll'}$ has a different constant value $v$.
\vspace*{-1.2cm}
\begin{figure}
\hspace*{-1cm}\epsfig{figure=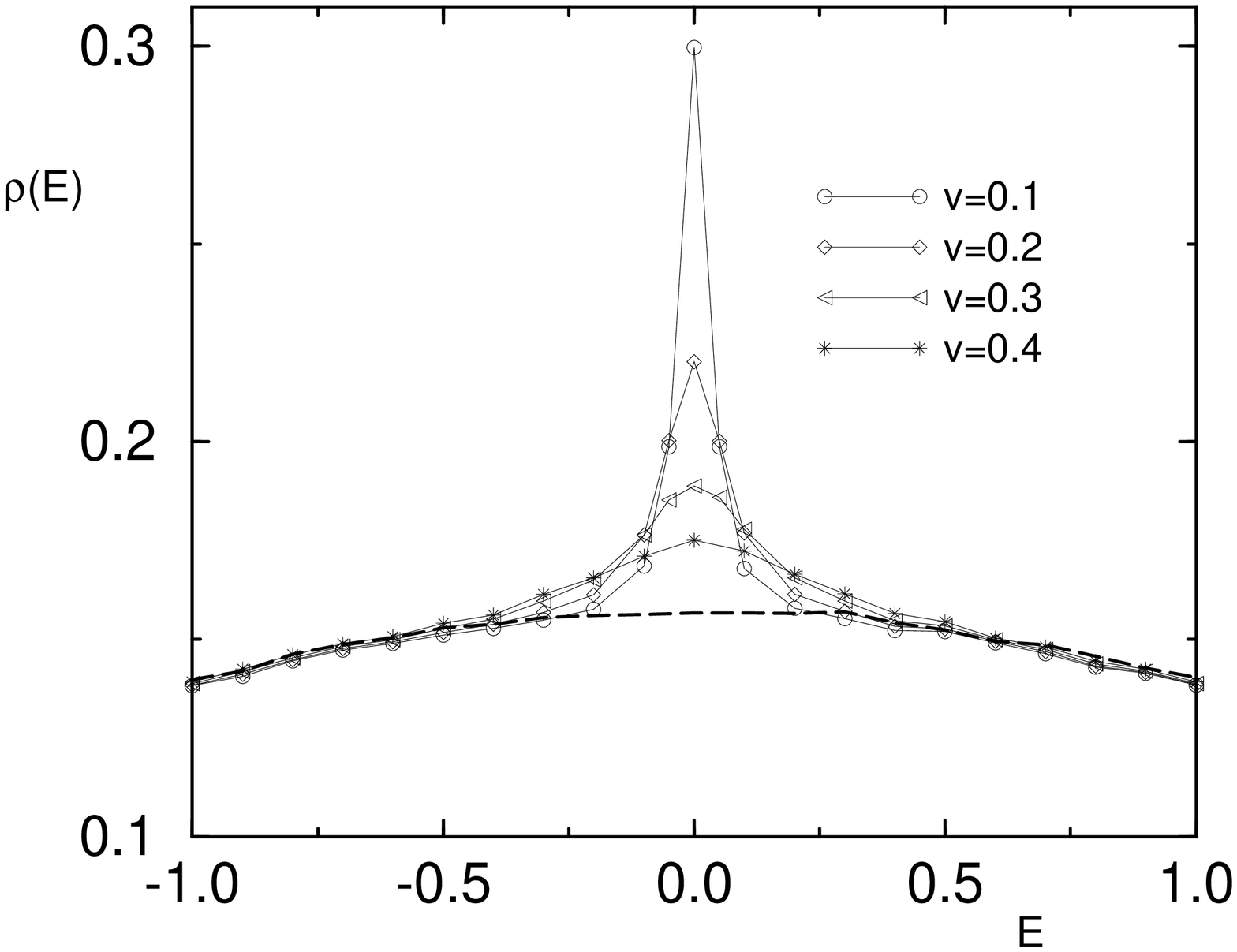,height=10cm,width=7cm,angle=270}
\noindent
{\footnotesize {\bf FIG. 1.}
The density of states $\rho(E)$ for the 2d--Anderson model in
the presence of resonance states with energy $E_r=0$ and various values of the
coupling constant $v$ corresponding to different symbols. The thick dashed line
corresponds to the Anderson model without resonances. In all cases the disorder
strength is $W=4$.}
\end{figure}

To evaluate the density of states $\rho(E)$ of this system, we start by writing
the original $(N+m)\times (N+m)$ Hamiltonian matrix $H$ in the following form
\begin{equation}
H=\left( \begin{array}{cc} \begin{array}[b]{c}
H_0 \end{array} & V \\
V^T & E_r\times I_m \end{array}
\right) .
\label{ham}
\end{equation}
This form can be obtained easily by separating the resonant states from
random states. $V^T$ denotes the transpose of $V$ and $H_0$ is a $N$--
dimensional matrix that describes the part of the Hamiltonian
$H$ without resonances. Here $I_m$ is the $m$--dimensional unit matrix, and
$E_r$ is the resonance energy of each of the $m$ resonant scatterers. The
rectangular matrix $V$ couples the $m$ resonances to $H_0$. All the matrix
elements of each column of $V$ are zero except four with value $v$.\\
The Green's function corresponding to the Hamiltonian (\ref{ham}),
can be written as
\begin{eqnarray}
\label{green1}
G&=&\left( \begin{array}{cc} \begin{array}[b]{c}
G_0 \end{array} & G_V \\
G_V^T & G_r \end{array}
\right) \nonumber \\
&=&\left( \begin{array}{cc} \begin{array}[b]{c}
E+i\eta-H_0 \end{array} & -V \\
-V^T & (E-E_r)\times I_m +i\eta\end{array}
\right)^{-1}.
\end{eqnarray}
To determine the different matrices in $G$, we use the previous relation to
obtain
the following set of equations:
\begin{eqnarray}
(E-H_0+i\eta)G_V-VG_r=0\,\,\,,\nonumber \\
-V^TG_V+[(E-E_r)I+i\eta]G_r=1.
\label{green2}
\end{eqnarray}
Solving (\ref{green2}) with respect to $G_r$ we get the expression
\begin{equation}
G_r=\frac{1}{ (E-E_r)I+i\eta-V^TG_aV}
\end{equation}
where $G_a=\frac{1}{E-H_0+i\eta}$. In the limit of weak coupling $v\ll 1$
and low concentration of resonances one can make the approximation $G_a
\simeq G_0$. Then the density of states is given by
\begin{equation}
\rho(E)= -\frac{1}{i\pi}\left({\it T}r G_0 +{\it T}r\frac{1}{(E-E_r)I+
i\eta-V^TG_0V}\right)
\label{green3}
\end{equation}
which, in the limit of $v\ll 1$ and low concentration of resonances, can be
written as
\begin{equation}
\rho (E) = \rho_0(E) + m\frac{\Gamma/2\pi} {(E-E_r)^2+(\frac{\Gamma}{2})^2} ,
\label{dos}
\end{equation}
where $\rho_0(E)=-\frac{1}{i\pi}{\it T}r G_0 $. To a good approximation,
$\rho_0(E)$ can be identified as the density of states of the standard
two--dimensional Anderson model without resonances. The effect of the
resonances is given by the second term on the right hand side of
Eq.~(\ref{dos}).
It exhibits a resonance enhancement near the resonance energy $E_r$ described
by a Lorentzian peak. The width, $\Gamma \approx 2\pi \rho_0 v^2$, is given
by the decay width of individual resonances. Based on physical grounds, we
expect that in the limit of weak coupling $(v \ll 1)$ there will be a strong
degeneracy due to the resonance states. This degeneracy is responsible for
the resonance
enhancement of the DOS at $E=E_r$. Gradually, as the coupling with the other
states is increased, the degeneracy and consequently the peak in $\rho(E_r)$
decrease. These expectations are in full agreement with Eq.~(\ref{dos}).
The same result has been found in the extended RMT, Ref. \cite{EKW},
with the use of the super-symmetry formalism. Here however Eq.~(\ref{dos})
appears natural under the frame of the tight-binding picture.
\vspace*{-1.2cm}
\begin{figure}
\hspace*{-1cm}\epsfig{figure=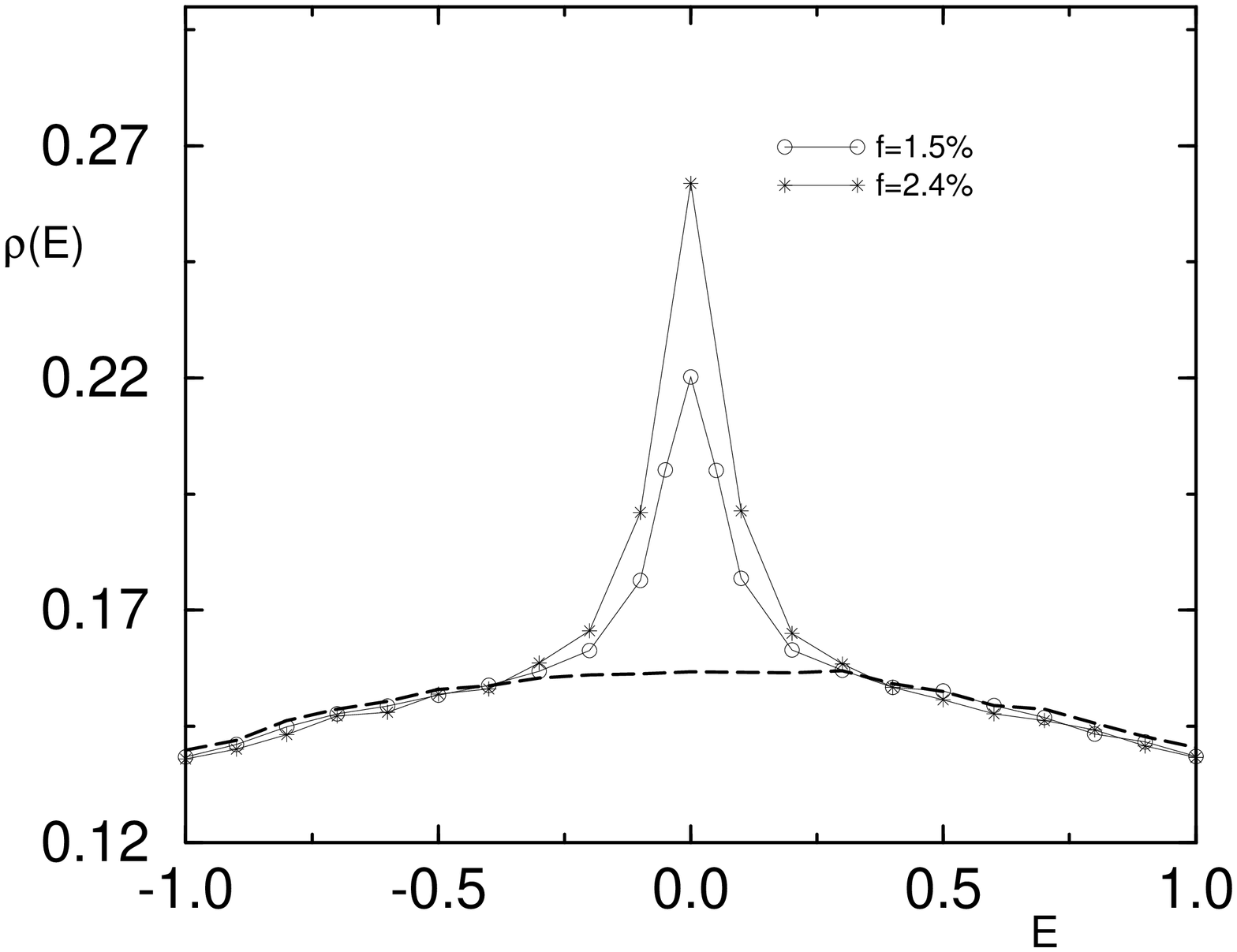,height=10cm,width=7cm,angle=270}
\noindent
{\footnotesize {\bf FIG. 2.}
Density of states $\rho(E)$
for the 2d--Anderson model in the presence of resonances with resonance energy
$E_r=0$, coupling constant $v=0.2$ and various concentrations. The thick dashed
line corresponds to the Anderson model without resonances. In all cases the
disorder strength is $W=4$.}
\end{figure}

To verify the validity of the approximations used in the derivation of
Eq.~(\ref{dos}) and to go beyond the weak coupling limit, we have performed numerical simulations of the 2d--Anderson
model with $m$ randomly distributed resonant states. Our main concern in this
study is to investigate how the DOS and the localization length behave as a
function of energy for various resonance concentration $f=m/(N+m)$ and
coupling strengths $v$. To evaluate these quantities, we used the iterative
procedure developed in \cite{MK83}. This method is particularly suitable for
the calculation of various quantities for a macroscopically large system
described by a microscopic Hamiltonian. The system we considered is a cylinder
of width $M$ in $y$ direction and length $L$ in $x$ direction. For $L$
large compared to $M$ the sample is essentially a quasi--one--dimensional
system. For the above calculations, we have used samples of length $L= 10^5$
and width up to $M=100$.

We first examine the behavior of $\rho(E)$ as a function of the coupling
constant
$v$ and the resonance concentration $f$. Our numerical results for low
resonance
concentration and various values of the coupling $v$ are reported in Fig.~1.
As we can see, at $E=E_r$ the DOS shows a resonance enhancement which is very
well reproduced by the analytical formula (\ref{dos}). As the resonance
coupling
$v$ increases we observe a gradual decrease of the hight of the Lorentzian peak
as well as an increase of its width. In the limit $v=1$ we recover the Anderson
case. We also investigated the dependence of the DOS on the concentration $f$
of
resonances in the disordered sample. In the weak coupling limit, we found that by increasing the
concentration $f$
the Lorentzian peak at the DOS increases in high (see Fig.~2) in 
agreement
with the analytical formula (\ref{dos}). Our numerical calculations, for small
resonance coupling $v=0.2$, are presented in Fig.~2. for two particular values of concentration.
We remind that exactly the same behavior of $\rho(E)$ was observed for the
one-dimensional case \cite{EK99}.
\vspace*{-1.2cm}
\begin{figure}
\hspace*{-1cm}\epsfig{figure=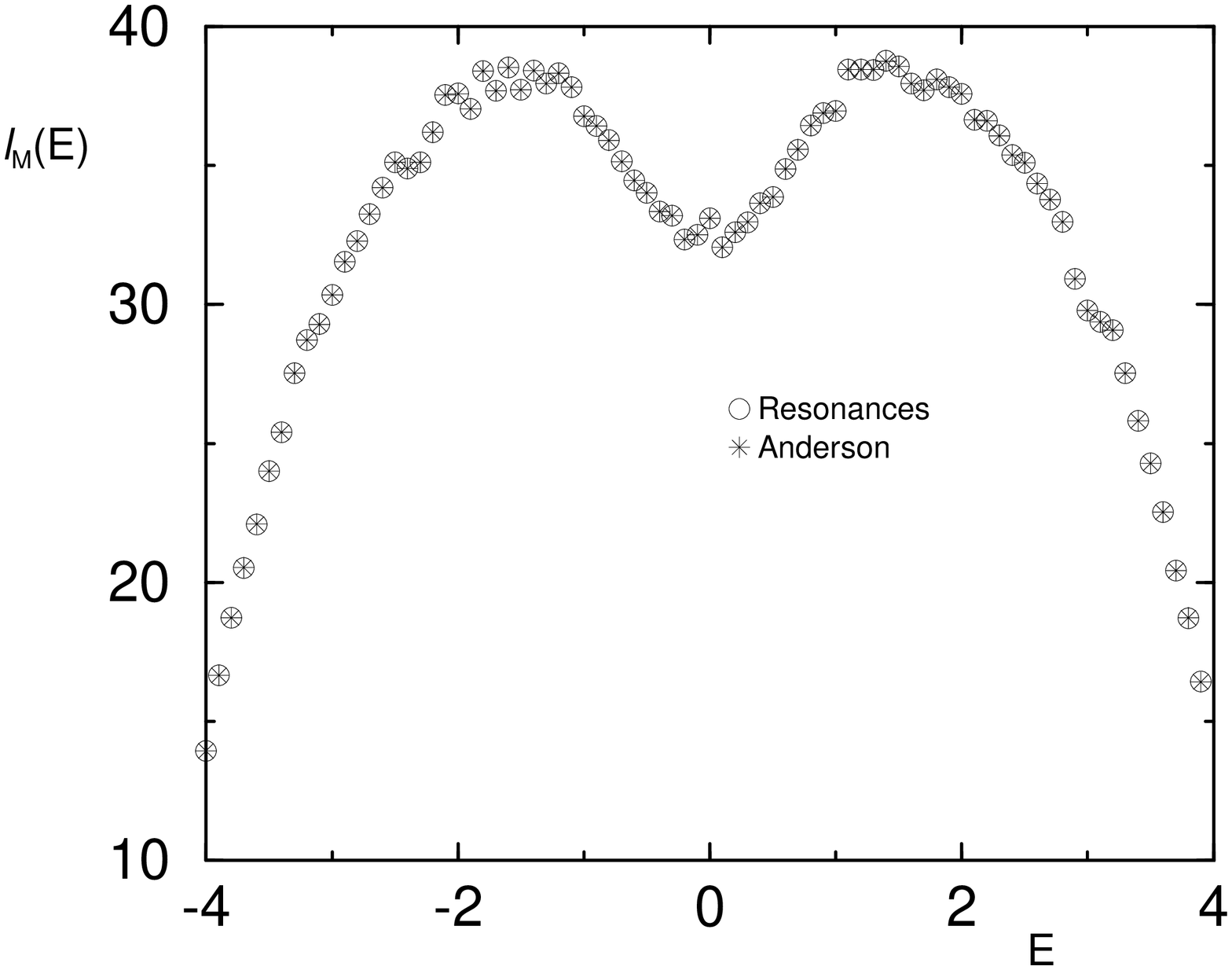,height=10cm,width=7cm,angle=270}
\noindent
{\footnotesize {\bf FIG. 3.}
Localization length $l_M(E)$ in the presence ($\circ$) and absence
($\ast$) of resonances calculated numerically for the 2d--Anderson model as a
function of energy. The width of the strip is $M=20$, and the disorder strength
is $W=4$. The resonance energy is $E_r=0$, while the resonance concentration
and coupling are $f=2.4$ and $v=0.2$ respectively. It is clear that $l_M(E)$
is not affected by the presence of resonances. }
\end{figure}

Next, we calculate the localization length $l_{\infty}(W,E)$, needed for the
determination of diffusion constant $D(E)$. Using the same numerical approach
\cite{MK83} we define a localization length $l_{M}(W,E)$ depending on $M$,
$W$ and $E$. The asymptotic localization length $l_{\infty}(W,E)$
 corresponding to the
infinite system, is then defined in the limit of $M\rightarrow \infty$ i.e.
\begin{equation}
\label{inflocle}
l_{\infty}(W,E)\equiv l_{M\rightarrow\infty}(W,E)
\end{equation}
The numerical results, performed for different values of resonance 
concentration, show that $l_{M}(E)$ can be considered
approximately independent from the concentration $f$ and from the resonance coupling
$v$ (as long as the concentration is not extremely large). Our results remain
 practically the same for all values of $M$ we have considered in our calculations (up to
$M=100$) and $f$. In Fig.~3 we present the behavior of the localization length $l_M(E)$ , for one particular value of $f$, vs. the corresponding one of the
standard Anderson model.
This behavior of the localization length is totally different from the one
found
in the one-dimensional case where $l_{\infty}(E)$ decreases drastically
everywhere except close to the resonance $E_r$ where it is approximately equal
to the corresponding one obtained for the standard Anderson model \cite{EK99}.
The underline physical reason is that the electron in the two-dimensional
geometry can bypass the resonances following a different path. We expect that
the same behavior will appear also in three-dimensions. In contrast, in the
one-dimensional case the electron has to pass through the resonance and
thus to suffer resonance tunneling. Thus, the one dimensional structure acts
as a filter allowing only electrons with energy $E_r$ to transmit.

We now turn our attention to the calculation of the diffusion constant which will be the
linking observable with the experimental results \cite{ATLT91}.   In the
quasi-one
dimensional multichannel case the diffusive constant $D(E)$ is related with
$l_{\infty}(E)$ and $\rho(E)$ through the relation \cite{efe}
\begin{equation}
\label{difco}
D(E) =\frac {l_{\infty}(E)} {4\pi\rho(E)}.
\end{equation}
Then, it becomes clear from the previous analysis that $D(E)$ will show a dip
near the resonance energy $E_r$ (see also \cite{LL96}), due to the resonance
enhancement of the density of states $\rho(E)$.
However, in order to compare the behavior of $D(E)$ with the experimental
results \cite{ATLT91} we have in addition, to take into account the fact
that the strength of the resonance coupling $v$ depend on the concentration
of resonances $f$ as $v\sim f^{5/6}$ \cite{EKW}. Fig.~4 shows $D(E)$ vs.
energy for different concentrations. In agreement with the experimental
results \cite{ATLT91}, we find that an increase of $f$ causes a decrease of $D$
far away from the resonance while the dip near the resonance energy $E_r$
widens and eventually disappears for $v\sim 1$. This result identifies the resonant behavior of the DOS as the main
explanation for the dip in the diffusion constant. Moreover, it establishes
the equivalence of transport properties in the presence of resonances, between
classical and quantum waves.
\vspace*{-1.2cm}
\begin{figure}
\hspace*{-1cm}\epsfig{figure=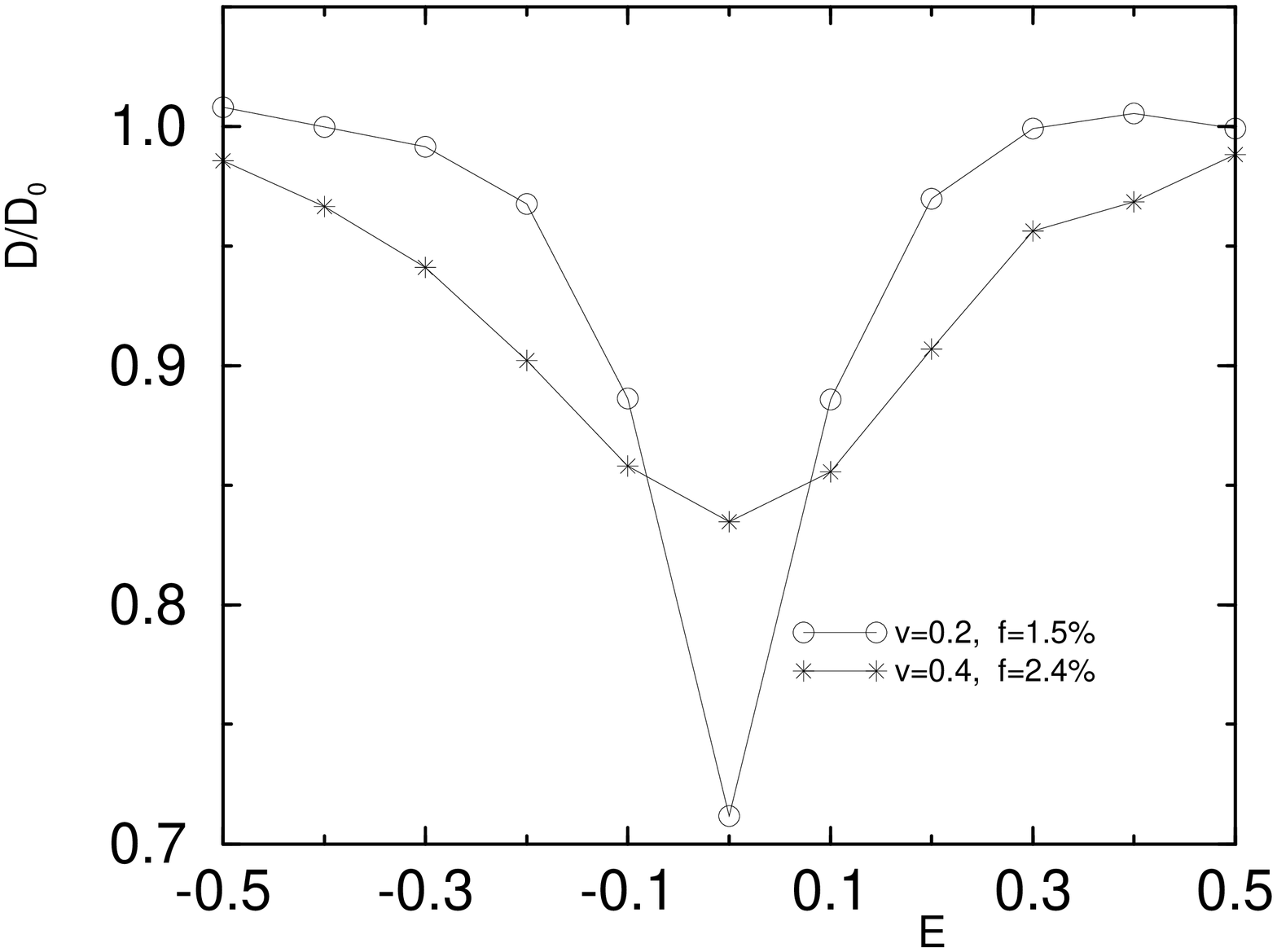,height=10cm,width=7cm,angle=270}
\noindent
{\footnotesize {\bf FIG. 4.}
Normalized diffusion constant  $D(E)/D_0(E)$ calculated numerically
for the 2D--Anderson model as a function of energy. The different curves
correspond to different concentrations $f$ of the resonances ($D_0$ is the
diffusion constant in the absence of resonances).}
\end{figure}

In conclusion, we have studied, both analytically and numerically, the effect
of
resonances on the transport properties of electrons in two--dimensional
tight--binding
Anderson model.  At low concentration and weak coupling the DOS is affected
strongly
by exhibiting a Lorentzian peak at resonance energy. As the coupling strength
increases,
the width $\Gamma$ of the Lorentzian increases and the resonance structure in
DOS
is smeared out. Contrary, the localization length is unaffected from the
resonances.
An application of the formula, $l_\infty(E)=4\pi\rho(E) D(E)$ \cite{efe},
identifies
the behavior of $\rho(E)$ as the explanation for the observed experimental
results
of the diffusion constant. This establishes the analogy between classical and
quantum waves. Our results are similar to those obtained within the ERMT
\cite{EKW}.

{\it Acknowledgements} We appreciate discussion with Y. Imry and C. Tejedor. B. E  gratefully
acknowledge the support of the European Union through the Training and 
Mobility of Researchers Ultrafast Network .

\end{multicols}

\begin{thebibliography}{10}
\bibitem{ping} For a review, see {\it Scattering and Localization of Classical
Waves in Random Media}, edited by Ping Sheng (World Scientific, Singapore,
1990).

\bibitem{IMRY} Y. Imry, {\it Introduction to Mesoscopic Physics}, (Oxford
University Press, New York, 1997).

\bibitem{GGW97} T. Guhr, A. Muller-Groeling, H. Weidenm\"uller, Phys. Reports
{\bf 299} (1998) 189.

\bibitem{ATLT91} Van Albada M., Van Tiggelen B. A., Lagendijk A. and Tip A.,
Phys. Rev. Lett. {\bf 66} (1991); see also N. Garcia, A. Z. Genack, and A. A.
Lisyansky, Phys. Rev. B {\bf 46}, 14475 (1992)

\bibitem{busch} C. M. Soukoulis, S.Datta, and E. N. Economou, Phys. Rev. B
{\bf 49}, 3800 (1994); K. Busch. C. M. Soukoulis and E. N. Economou,
Phys. Rev. B {\bf 50}, 93 (1994).

\bibitem{EKW} B. Elattari, V. Kagalovsky, and H.A. Weidenm\"uller, Phys. Rev.
B {\bf 57} 11258 (1998).

\bibitem{rmt}  see e.g. M. L. Mehta, {\em Random Matrices}, 2nd ed. (Academic
 Press, New York, 1991), and references therein.

\bibitem{EKW1} B. Elattari, V. Kagalovsky, and H.A. Weidenm\"uller, Phys.
Rev. E, {\bf 57} 2733 (1998); B. Elattari, V. Kagalovsky, and H.A.
Weidenm\"uller, Europhys.  lett., {\bf 42} 13 (1998).

\bibitem{EK99} Brahim Elattari and Tsampikos Kottos, Phys. Rev B {\bf 59},
R5265 (1999).

\bibitem{MK83} A. MacKinnon, B. Kramer, Z. Phys. B-Condensed Matter {\bf 53},
1 (1983); A. MacKinnon, Z. Phys. B-Condensed Matter {\bf 59}, 385 (1985).

\bibitem{efe} K. B. Efetov, Adv. Phys. {\bf 32}, 53 (1983).

\bibitem{LL96} D. Livdan and A. A. Lisyansky, Phys. Rev. B {\bf 53}, 14843
(1983).
\end{thebibliography}
\end{document}